\documentstyle[11pt,mrs2003,epsfig]{article}
\begin{document} 
\title{THE GraS HYPOTHESIS: A MODEL FOR DARK MATTER-BARYONS GRAVITATIONAL INTERACTION}

\author{FEDERICO PIAZZA$^{1,2}$ and CHRISTIAN MARINONI$^3$\\}

\affil{$^1$Dip. di Fisica, Universit\`a di Milano Bicocca, P.zza delle Scienze 3, 20126 Milan, Italy\\
$^{2}$ Institute of Cosmology and Gravitation, University of Portsmouth, Portsmouth PO1 2EG, UK\\ 
 $^3$ Laboratoire d'Astrophysique de Marseille,  Traverse du Siphon B.P.8, 13376 
Marseille, France}

\begin{abstract} 
According to the gravitational suppression (GraS) hypothesis
the gravitational interaction between exotic (dark) and 
standard (visible) matter is suppressed below the kpc scales. 
We review the phenomenological motivations at the basis of this idea and its 
formal implementation by means of a Yukawa contribution to the usual Newtonian potential. 
We also discuss a class of astrophysical phenomena that may help in testing  this scenario.

\end{abstract} 
 
\section{Introduction} 
The success of the cold dark matter (CDM) paradigm in explaining  a wide range 
of properties of the visible structures on scales $\geq 1$ Mpc \cite{PEA,BOP} does not extend down 
to sub-galactic scales. Some inconsistencies between 
observations and theory (via simulations)  still lack a satisfactory interpretation, 
 indeed    
a possible scenario of  ``CDM crisis'' on small galactic scales has been evocated \cite{MOORE}.

One issue is that, high-resolution observations of the inner  rotation curves of low-surface brightness (LSB)
galaxies suggest that the dark matter is  distributed in spherical halos with  nearly constant density cores 
\cite{MCG,BL2,MAR}, whereas DM simulations systematically predict  much steeper density profiles 
$\rho \sim  r^{-\beta}$ with $\beta \sim$ 1-1.5  (see, e.g. \cite{NFW,MOO,BUL}). 
 It now seems that the cuspy density profile of halos that form in simulations of the dark matter
component is a robust and reproducible feature of the collisionless dynamics of halo formation 
(e.g., \cite{POW}), and, as a matter of fact,
steep profiles consistent with the simulations have recently been ``seen'' through gravitational 
lensing at the center of galaxy clusters \cite{DAH}.
A parallel problem is that a 
scale-invariant primordial spectrum of perturbations generates
significantly more virialized objects of dwarf-galaxy mass 
than are observed around the Milky Way  \cite{KWG,KKV,MAT}.
Again, preliminary observations of multiple image gravitational lenses \cite{DK} seem to confirm   
the presence of these mass  substructures, thus indicating that a possible solution to this problem 
may be found in some mechanism that prevents star formation in most of the low mass galactic satellites.
Intriguingly enough,  these problems appear to be characterized by the same physical scale length of the
order of the kiloparsec. 

It is known that, in order to have efficient structure formation, dark matter (DM) must be sufficiently decoupled from 
the Standard Model particles. 
As a consequence, the direct detection of DM is (and is proving to be \cite{THIS}) a difficult task. Moreover,  DM 
properties must be inferred from the dynamic and distribution of the visible component, 
which in turn are affected  and shaped by DM only through gravitation. 
In what follows we review the recently proposed \cite{PM} gravitational suppression (GraS) hypothesis which addresses the above mentioned kpc-scale problems via a modification of the gravitational law
between DM and baryons. Of course, the theoretical framework 
of four dimensional General Relativity hardly accomplishes departures from the Newton law, unless other light field-mediated forces are introduced. 
Nevertheless, from an empirical stand point, a 
modification of gravity may be considered  as a ``minimal'' attempt to solve the above mentioned inconsistencies
between observations and theory, since it's just and only gravity that 
rules the reciprocal behavior of baryons and DM. In other terms we do not add or modify hypothetical 
properties of DM particles (e.g.. \cite{SPE,KAM})  but, applying "Occam razor",
we suggest a modification of a measurable phenomenon  whose general consequences are directly testable
and falsifiable.

As recalled by Tremaine \cite{TRE},  there have been two major observational  issues in the solar system
since 1800: the unexplained residuals in the orbit of Uranus that led to the discovery of Neptune, and the anomalous 
precession of Mercury perihelion that was explained by General Relativity. The {\em large scale} issue 
was explained by the discovery of new matter and the {\em small scale} anomaly by a radical new physical law.
In analogy, the dynamics of spiral galaxies at {\em large radii} has lead to the discover of dark matter, while 
the {\em inner} rotational behavior of low surface brightness (LSB) galaxies may contain a hint of fundamental new physics. 
We leave speculations about the theoretical origin
of such a proposed modification to future work. At present we adopt  an instrumentalist view
and try to asses the performances of GraS as a working hypothesis.

\section{The Gravitational Suppression Hypothesis}

The GraS hypothesis has been formulated in \cite{PM} as follows: \emph{
In the Newtonian limit of approximation 
(small curvatures and small velocities in Planck units), the gravitational interaction between baryonic 
and non baryonic particles is suppressed on small ($\sim $ kpc) scales.} 
The picture is very simple and schematically represented in Fig. 1: nothing changes in each of the two sectors
(visible-dark) with respect to Newtonian gravity but, as two particles (one ``dark'' and one ``visible'') get closer 
than a kpc to each other, they experience a suppression of the usual gravitational Newton law. 

In order to model a general small scale modification of gravity we have used a Yukawa-type correction to the usual $1/r$ 
law. The resulting gravitational potential between two point-like particles of different nature goes like
\begin{equation} \label{1}
\phi(r) \, \propto \, -\,  \frac{1}{r}(1+ \alpha e^{-r/\lambda}) \, ,
\end{equation}
where $r$ is the distance between the two particles. The parameter $\alpha$ gives the strength 
and the type (repulsive or attractive) of the Yukawa contribution. As $\alpha$ gets close to $-1$ the total gravitational
interaction tends to be \emph{totally suppressed} in the small scale limit. The length $\lambda$, on the other hand, gives
the typical scales over which such a suppression is effective. 
Accidentally, we note that the same kind of interaction is generated by a light scalar mediated force. However,
a universal scalar interaction cannot be active only in the $mixed$ sector without affecting also the dynamics 
internal to each sector. 

%
%
\begin{figure} 
\vspace*{1.25cm}  
\begin{center}
\epsfig{figure=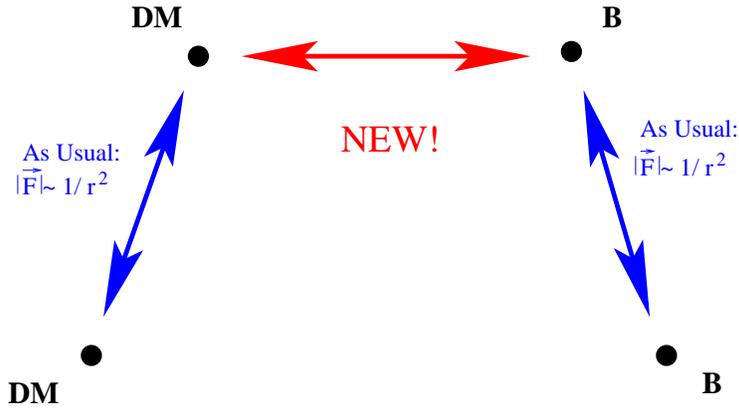,width=10cm}  
\end{center}
\vspace*{0.25cm}  
\caption{A schematic representation of the GraS hypothesis: the suppression of gravity on kpc scales is active only
between ``dark'' and ``visible'' particles. The Newtonian dynamics inside each of the two sector is untouched.} 
\end{figure}

\section{Fitting the parameters} 
 
Low surface brightness (LSB) galaxies are very good  laboratories to test our model.
Since they are dominated by dark matter \cite{BLO} we can neglect the dynamical contribution 
of the disc to the velocity rotation curves. Moreover, since GraS doesn't change the usual Newton law between DM particles,  the general predictions of numerical simulations concerning the density profiles of DM halos hold. In particular, we expect inner
power-law cusps of the type $\rho \sim  r^{-\beta}$ with $\beta \sim$ 1-1.5 \cite{NFW,MOO,BUL} .
In \cite{PM} the velocity rotation curve generated by a
power-law density profile of generic slope $\beta$ has been calculated within the framework of 
the Yukawa model of equation (\ref{1}). Such a theoretical curve has been compared to  
a sample of high resolution LSB rotation curves measured and reduced by different authors
\cite{BL2, MAR, BLA, BSL} in order to fix  the universal  parameters of our model, $\alpha$ and $\lambda$, and to
find the best value $\beta$ of the internal slope consistent with the GraS hypothesis (see Fig. 1 in \cite{PM}).

%
%
\begin{figure} 
\vspace*{1.25cm}  
\begin{center}
\epsfig{figure=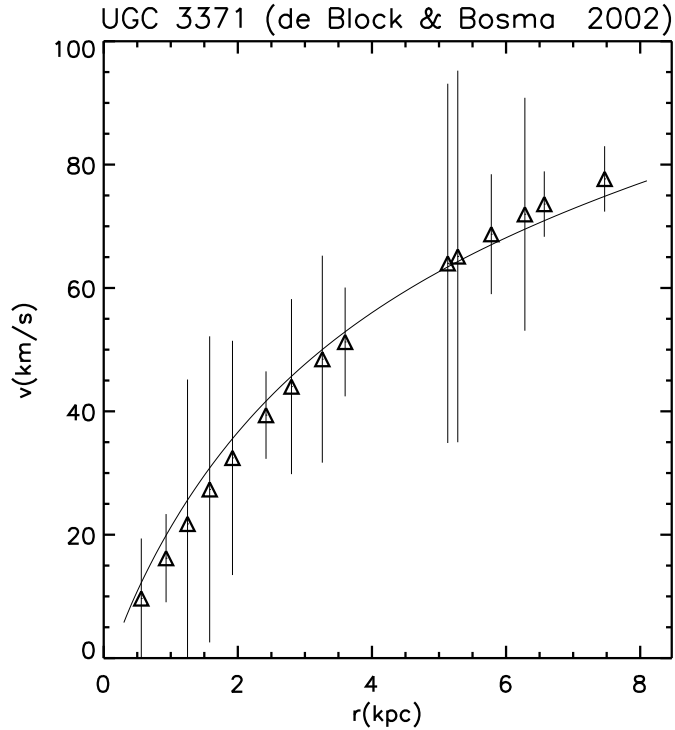,width=10cm} 
\end{center}
\vspace*{0.25cm}  
\caption{  
The universal velocity  model derived in \cite{PM}   using the GraS potential and 
assuming no dynamical contribution from the baryon in the disk (minimal disk hypothesis)
is superimposed to rotational data of the LSB galaxy 
UGC3371 \cite{BL2}. Errorbars represent 1$\sigma$ uncertainties. } 
\end{figure} 

The value of $\alpha$ ($\alpha = -1 \pm 0.1$) is remarkably stable in all the galaxies considered and seems to point
towards a \emph{total} suppression scenario where the gravitational attraction between two particles of different 
nature is really ``cut-off'' at 
small distances. The characteristic scale below which  the suppression mechanism is active  
is constrained to be  $\lambda = 1.1 \pm 0.08$ kpc. 
Finally, the inner density slope, $\beta = 1.35 \pm 0.05$, is in good agreement with the
simulations \cite{NFW,MOO,BUL} and in particular with some recent high-precision ones which seem to indicate the value  $\beta = 1.2$ \cite{POW}.
The rotational velocity predicted by such a set of universal parameters  
has been superimposed  in Fig 2. to the observed velocity curve  of a new LSB galaxy \cite{BL2} for which 
sufficiently high resolution inner rotation data  (but unfortunately large observational errors for a
meaningful fit) are available.

\begin{figure} 
\vspace*{1.25cm}  
\begin{center}
\epsfig{figure=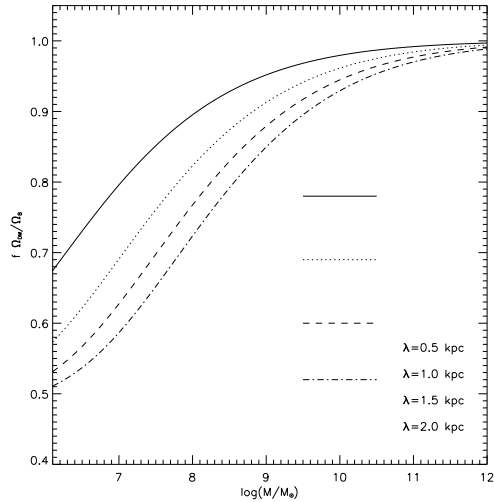,width=7cm}  
\end{center}
\vspace*{0.25cm}  
\caption{Baryon fraction at turn-around epoch   in units of its mean cosmological value $\frac{\Omega_b}{\Omega_{m}}$ as a function
of the mass of the collapsed object an for different values of $\Lambda$.} 
\end{figure}

\section{GraS: Predictions}

GraS provides a cosmological mechanism that segregates the fraction of baryons in halos of different mass 
sizes. The reason is rather intuitive: small mass halos of sizes up to a kpc are much less 
effective in attracting the 
cosmologically diffused baryons. It may be worth noting that dwarf galaxies of $10^9$ M$_{\odot}$ 
typically grew out of fluctuations which, at decoupling, were not larger than $10^{-1}$ kpc. 
In \cite{PM} this effect has been calculated at linear order in perturbation theory considering
the baryon fraction 
\begin{equation} \label{2}
f\ = \ \frac{\rho_B}{\rho_D}\ = \ (\Omega_B / \Omega_D)\, \frac{1 + \delta_B}{1+ \delta_D}
\end{equation}
at the epoch of turnaround. As a result, the baryon fraction in halos of $10^6$ solar masses is 
about half the average cosmological value $\Omega_B / \Omega_D$ (see Fig. 3). This linear approach, however, is likely to 
underestimate the GraS effects. At turnaround, in fact, where the linear theory approximation  still holds, 
 $\delta_D \simeq 1.1$ and, 
even for completely frozen baryonic fluctuations $\delta_B \simeq 0$, the fraction on the RHS
of (\ref{2}) cannot be lower than about $1/2$. 
As a consequence galaxy formation is not 100\% efficient in trapping 
baryons inside small halos, and   a relevant fraction 
of the total baryonic budget should consist of cool baryonic clouds 
outside virialized structures. It is tempting to speculate on how this mechanisms   
may help in interpreting the fact that  part  of the baryons expected in standard nucleosynthesis models 
are yet undetected in the present day universe.

Moreover, loosening the tight coupling between baryons and exotic particles, 
other complementary astrophysical mechanisms such as Supernovae feedback 
and photoionization may become  more effective in preventing star formation. 
For instance, the GraS predicted binding energy of 
baryons in halos of $10^8$ solar masses is about a factor of one tenth less than predicted by Newtonian 
gravity.  A more realistic picture of gas/star density distribution and evolution in small 
galaxies could be obtained by implementing GraS non-linear, non-gravitational  dynamical effects
into high resolution hydrodynamical simulations.

Simulations may also help in assessing the effect of GasS on the dynamical coupling between the stellar bars in 
galaxy discs and the surrounding dark matter halo and the subsequent evolution of both components.
One may speculate that the apparent absence of slow bars in galaxies is due to a suppression 
of the dynamical friction between the baryons and CDM particles and thus to a less effective 
transfer of angular momentum between the bar and halo components. 

In summary we have argued that the severe challenges facing DM models may  indicate that a serious
examination of the way baryons respond to a dark matter potential on small scales should be 
considered.

\acknowledgements{In Marseille, during the days of the conference, we've enjoyed the good company of, among others, 
Latham Boyle, Marco Bruni, Justin Khoury, Roberto Mainini, Roger Malina, Saul Perlmutter, Claudia Quercellini and
Matthew Parry. It is a pleasure to thank all these people for stimulating discussions about this and 
other subjects}

\vfill 
\end{document}